\documentclass[12pt,preprint]{aastex}

\newcommand{\kms}{km s$^{-1}\,$}
\newcommand{\tu}{2s$^2$2p$^3$ }
\newcommand{\fo}{[\ion{O}{2}] }
\newcommand{\oz}{O$^+$}
\newcommand{\Dd}[1]{$^2\textrm{D}^o_{#1}$}
\newcommand{\Pp}[1]{$^2\textrm{P}^o_{#1}$}
\newcommand{\Ss}{$^4\textrm{S}^o_{3/2}$}
\newcommand{\Ddn}{$^2$D$^o$}
\newcommand{\Ppn}{$^2$P$^o$}
\newcommand{\Ssn}{$^4$S$^o$}
\newcommand{\Ddx}{^2\textrm{D}^o}
\newcommand{\Ppx}{^2\textrm{P}^o}
\newcommand{\Ssx}{^4\textrm{S}^o}

\newcommand{\oh}{OH }
\newcommand{\cmm}{cm$^{-1}$}
\newcommand{\cmc}{cm$^{-3}$}
\newcommand{\lam}{$\lambda$}
\newcommand{\lamm}{$\lambda\lambda$}

\shorttitle{Doublet O II}
\shortauthors{SHARPEE ET AL.}
\begin{document}

\title{Measurements of the singly ionized oxygen auroral doublet lines
$\lambda\lambda$7320,7330\AA\ using high resolution sky spectra}

\author{Brian D. Sharpee, Tom G. Slanger, David L. Huestis, Philip
C. Cosby} \affil{Molecular Physics Laboratory, SRI International, 333
Ravenswood Ave, Menlo Park, CA 94025 } \email{brian.sharpee@sri.com,
tom.slanger@sri.com, david.huestis@sri.com, philip.cosby@sri.com}

\begin{abstract} 
The wavelengths of the individual \fo \tu \Dd{5/2} -- \tu \Pp{1/2,3/2}
\lam7320\AA\ and \fo \tu \Dd{3/2} -- \tu \Pp{1/2,3/2} \lam7330\AA\
auroral doublet component lines have been measured directly in the
nightglow for the first time, from high resolution spectra obtained
with the HIRES spectrograph on the Keck I telescope at the W.M. Keck
Observatory.  Specifically, we find 7320\AA\ and 7330\AA\ doublet
splittings of 1.077$\pm$0.003\AA\ and 1.080$\pm$0.003\AA\
respectively, with the former significantly larger than the often
quoted and utilized value of 0.8\AA\ from Moore (1949, Atomic Energy
Levels, Vol 1), and in line with NIST (1.07\AA) as well as more recent
astrophysical observations of the lines in planetary nebulae including
1.07\AA\ from DeRobertis, Osterbrock, \& McKee (1985,ApJ,293,459) and
1.09\AA\ from Barnett \& McKeith (1988,MNRAS,234,241).  Our results
suggest, however, that adjustments of +0.124\AA\ and +0.131\AA\ should
be made to current NIST wavelengths for the blue and red components of
the 7320\AA\ doublet respectively, while the wavelengths of the
7330\AA\ doublet components are little changed from current NIST
values.  The observed intensity ratio of 7320\AA/7330\AA\ from these
measurements agrees with the theoretical value calculated under
conditions of thermally populated $^2$P$^o$ levels.
\end{abstract}


\section{Introduction} \label{intro}

The auroral \fo \tu \Dd{5/2} -- \tu \Pp{1/2,3/2} \lam7320\AA\ and \fo
\tu \Dd{3/2} -- \tu \Pp{1/2,3/2} \lam7330\AA\ emission line doublets
(hereafter referred to as ``7320,7330'') are readily visible in the
spectra of low ionization astrophysical plasmas such as planetary
nebulae \citep{B55,B60,D85,S03} and \ion{H}{2} regions
\citep{B00,K03}, where the levels of the \oz(\Ppn) term are populated
by collisions between thermal electrons and \oz\ ions in the \Ssn\
ground term \citep{O89}.  As such, their strengths can be used to
determine the electron density in the moderate density regime
($10^5$--$10^{6}$ \cmc) of such objects \citep{BK88}, and are commonly
employed in combination with the nebular \fo \tu \Ssn -- \tu
\Dd{3/2,5/2} \lamm3726.032,3728.815\AA\ lines (hereafter
``3726,3729'') to measure electron temperature \citep{SO57,K99}.  The
doublets are also an important component of the airglow and aurora,
where \oz(\Ppn) is produced primarily by photoionization of atomic O
by solar EUV ($\lambda<666$\AA) and electron impact ionization (with
energy $<1$keV), respectively.  Their altitude-dependent emission
strength is used to calculate the quenching rate of \oz(\Ppn) ions
through collisions with N$_2$ and atomic O \citep{R77,C93,ST03}, the
altitude density profile of \oz(\Ppn) and photoionization rate of
atomic O \citep{M91}, and the ion convection velocity and \oz\ gas
temperature in the ionospheric F-layer \citep{S82,C03}.  The
intensities and line widths of the 7320,7330 doublet components have
also been employed to detect the existence and characteristics of a
permanent ``hot'' atomic oxygen corona existing above an altitude of
550 km \citep{Y80}.

We believe two recent studies demonstrate the necessity of determining
and publicizing accurate energies for the \oz\ \tu levels,
particularly the wavelengths and splitting of the 7320,7330 doublet
components.  In their imaging Fabry-Perot interferometric auroral
data, \citet{C03} measured a 7320 doublet separation of 0.796\AA\ from
their fit to the separable unresolvable, combined profile of their
observed 7320 component lines, close to the 0.8\AA\ (1.5 \cmm) which
they quote from \citet{S82}.  However, neither study could completely
resolve the separate lines because of their large intensity
difference.  Moreover, recent \textit{ab initio} and astrophysically
observed calculations of the splitting argue for a significantly
larger separation (see Table~\ref{table1}).  Indeed, the newer values
of the separation are larger than the 0.8\AA\ quoted from \citet{S82}
and measured by \citet{C03} by the order of the typical 0.1--0.8 \kms
ion-drift velocities targeted for measurement by both studies.  A
larger 7320 separation also calls into question both the \citet{C03}
interpretation of their observed line profile as the actual 7320 pair,
and the accuracy of their F-layer temperature determination, made from
a line profile fit in which the separation itself was a free
parameter.  Meanwhile, in their recent high resolution spectra of the
Orion Nebula, \citet{B00} found that the difference between their
observed 3726,3729 and tabulated (NIST) wavelengths, after correction
for nebular proper motion, is in good agreement with the same for the
7330 components, but is systemically larger (by $\approx$6
\kms=0.15\AA) than the same for the 7320 components, even though all
lines should originate in the same portion of the nebula.  This
discrepancy is a significant fraction of the total range of
ionization-energy dependent velocities seen in the Orion Nebula
outflow \citep{B00}.

High resolution spectroscopy of the night sky has been shown to be a
valuable tool for direct measurements of the line rest wavelengths
\citep{SO00}.  Presented here are what we believe to be the first
direct measurements in the airglow of the 7320,7330 doublet component
wavelengths, in which the doublets can be cleanly resolved into their
component lines.  These measurements have been used to re-evaluate the
energy levels of the \Ddn\ and \Ppn\ terms, with the aim of improving
their accuracy for use in the types of studies involving precision
measurements of drift and nebular velocities, such as those mentioned
above.

\section{Observations and Reductions} \label{obs}

Sky spectra were drawn from among those to be archived in the proposed
National Virtual Aeronomical Observatory
(NVAO)\footnote{\url{http://www.nvao.org}} \citep{H02} showing the
most prominent 7320,7330 lines, which were clearly distinguishable
from and/or deemed clearly stronger than other neighboring telluric
emission.  These sky spectra were all obtained with the HIRES echelle
spectrograph \citep{V94} on the Keck I telescope at the W.M. Keck
Observatory, and were originally associated with numerous deep-sky
astronomical targets.  The nominal instrumental resolution was in all
cases 45,000 (R$\approx$6.7 \kms).  Only spectra which had both
doublets in the same order (echelle orders 48 or 49) were chosen in
order to minimize any wavelength and flux calibration errors that are
often associated with echelle spectra.  An observing journal is listed
in Table~\ref{table2}.

\begin{figure}
\plotone{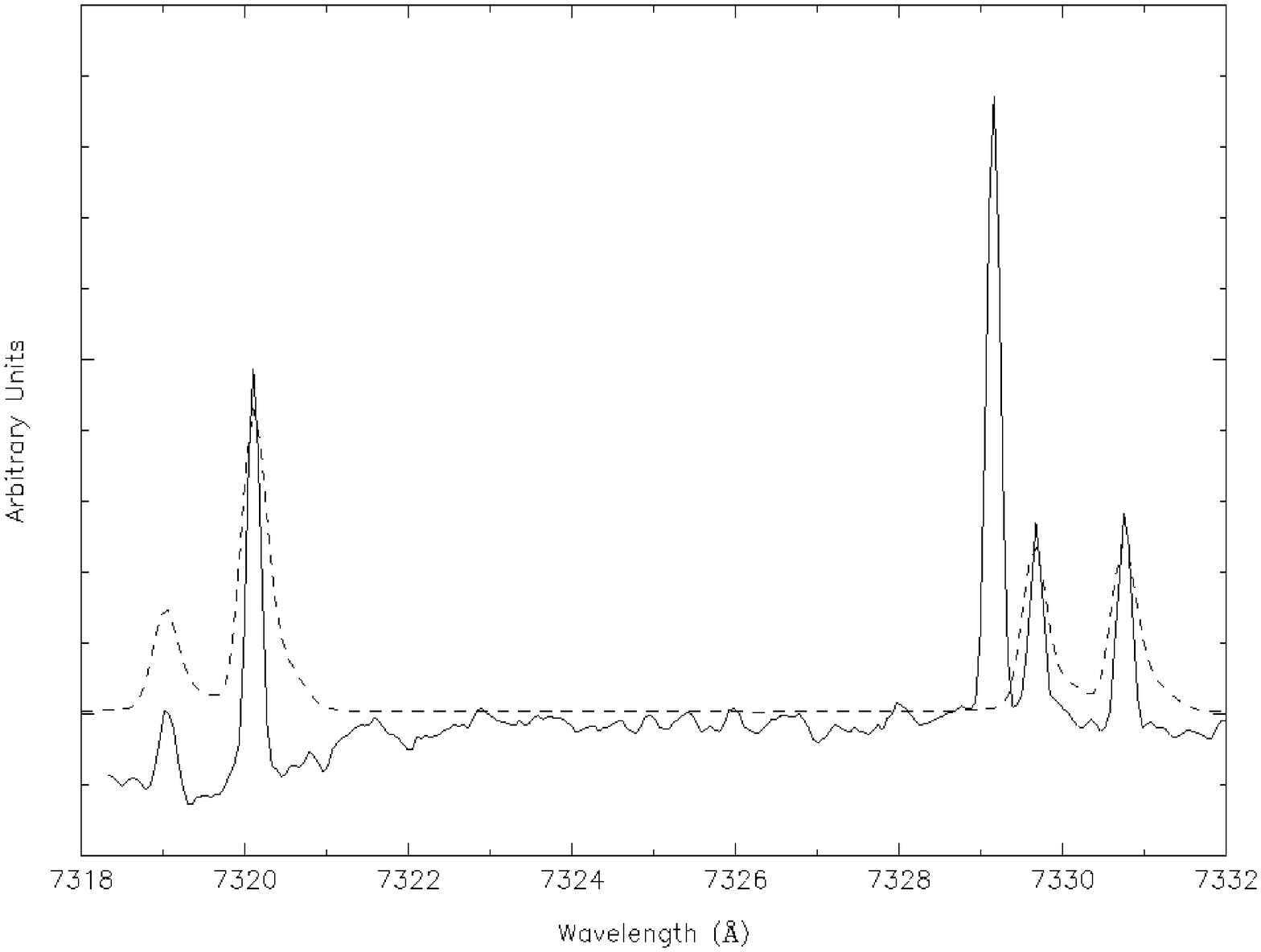}
\caption{A portion of a typical HIRES echelle spectrum showing the \fo
\tu \Ddn--\Ppn\ 7320,7330 doublet lines and \oh P$_{22}$(2.5) (8-3)
\lam7329.148 line.  The dashed line is a segment of the Orion Nebula
spectrum of \citet{B00}, scaled and shifted. \label{figure1}}
\end{figure}

All HIRES sky spectra were delivered to the NVAO archive, extracted
and reduced through wavelength calibration using the
MAKEE\footnote{\url{http://http://spider.ipac.caltech.edu/staff/tab/makee/}}
reduction routines developed by T.A.\ Barlow.  Line wavelengths were
determined by manually fitting a Gaussian and a straight line,
representing the local continuum, to each line profile.  Fluxes were
determined by summing the counts under the fitted function.  As can be
seen in Figure~\ref{figure1}, the high resolution of the HIRES spectra
completely separates the doublet components and neighboring \oh
telluric lines, and profiles were well-represented by Gaussian fits.

Besides the HIRES data, the tabulated 7320,7330 line wavelengths (in
air) from \citet{D85} (planetary nebula NGC 7027: R=100,000=3 \kms),
and 7320,7330 lines re-measured from the spectra of \citet{B00} (Orion
Nebula: R=30,000=10 \kms) and \citet{S03} (planetary nebula IC 418:
R=33,000=9 \kms), were used to supplement the HIRES data.
\citet{B60}, \citet{B00}, and \citet{S03} were also used for the
wavelengths of the 3726,3729 lines which are too weak to
be seen in the airglow.

\section{Calculations} \label{calc}

The tabulated wavelengths of the following \oh Meinel band (8-3) lines
from \citet{G98}: P$_{11}$(2.5) \lam7316.282, P$_{22}$(2.5)
\lam7329.148, P$_{11}$(3.5) \lam7340.885, and P$_{22}$(3.5)
\lam7358.667, if present in the same HIRES echelle order as 7320 and
7330, were compared to measured wavelengths and used to normalize each
HIRES spectrum to the same wavelength system.  A weighted average of
differences between the measured and tabulated wavelengths was
calculated for each HIRES data-set, with the above lines receiving
weights of (4,10,2,1) respectively, roughly proportional to their
relative theoretical intensities.  These corrections, ranging in
magnitude from 0.1--2.7 \kms (0.002--0.066\AA\ at 7325\AA), were
applied to all measured wavelengths in each HIRES spectrum.  The
statistical scatter between tabulated and observed wavelengths after
correction was 0.5 \kms ($\approx$0.01\AA\ at 7325\AA).

A linear chi-squared fit was then performed to simultaneously
determine the best set of shared 3726,3729 and 7320,7330 observed air
wavelengths for all data-sets in which the lines were contained, from
which a Doppler shift to correct for any residual internal kinematic
and proper motion could be determined and applied to all nebular data
(except for Bowen 1960, which are averages from numerous planetary
nebulae).  The minimized merit function took the form:
\begin{equation}
\label{eq1}
\chi^2 = \sum_i\sum_j w_j [{\lambda^i}_j(obs)-(\lambda^i(best)+\delta_j)]^2\,,
\end{equation}
where $i$ and $j$ are sums over all lines (3726,3729 and 7320,7330)
and data-sets respectively, $w_j$ are arbitrary weights given to each
data-set (see below), ${\lambda^i}_j(obs)$ are the wavelengths of each
line, if observed, in each data-set, $\lambda^i(best)$ the best shared
wavelength for each line, and $\delta_j$ the Doppler shifts (in \AA)
applicable to each nebular data-set.  In total 6 line wavelengths, one
for each 3726,3729 and 7320,7330 line, and 3 Doppler shifts, one for
each nebular data-set except for \citet{B60}, were determined in this
manner.  The weights $w_j$ (0.018 for Bowen, 0.16 for DeRobertis,
Osterbrock \& McKee, 1.00 for Baldwin et al.\ and Sharpee et al., and
10.00 for all HIRES data) were arbitrarily chosen to best represent
our perceived accuracy of the measurements involved, with nebular data
sets generally receiving lower values due to unknown amounts of
residual internal kinematic and proper motions.

Following conversion of all observed line wavelengths to vacuum
wave-numbers (cm$^{-1}$), linear chi-squared fits were made to
each level/term splitting value as combinations of those wave-numbers:
\begin{eqnarray}
\label{eq8}
\chi^2 & = & \sum_j w_j [\Delta(\Ddx-\Ssx) - \case{1}{2}(\nu_j(3726\textrm{\AA}) + \nu_j(3729\textrm{\AA}))]^2 \,, \\ 
\chi^2 & = & \sum_j \left\{w_j [\Delta(\Ppx-\Ddx) - \case{1}{2}(\nu_j(7319\textrm{\AA}) + \nu_j(7331\textrm{\AA}))]^2\right.  \nonumber \\
       &   & + w_j [\Delta(\Ppx-\Ddx) - \case{1}{2}(\nu_j(7320\textrm{\AA}) + \nu_j(7331\textrm{\AA}))]^2\left.\right\}  \,, \\   
 & & \nonumber \\ 
\chi^2 & = & \sum_j \left\{w_j [\Delta(\Ddx) - (\nu_j(3726\textrm{\AA}) - \nu_j(3729\textrm{\AA}))]^2 \right.\nonumber \\ 
       &   & + w_j [\Delta(\Ddx) - (\nu_j(7319\textrm{\AA}) - \nu_j(7330\textrm{\AA}))]^2 \nonumber \\
       &   & + w_j [\Delta(\Ddx) - (\nu_j(7320\textrm{\AA}) - \nu_j(7331\textrm{\AA}))]^2 \left.\right\}\,, \\  
 & & \nonumber \\ 
\chi^2 & = & \sum_j \left\{w_j [\Delta(\Ppx) - (\nu_j(7319\textrm{\AA}) - \nu_j(7320\textrm{\AA}))]^2 \right.\nonumber \\
       &   & + w_j [\Delta(\Ppx) - (\nu_j(7330\textrm{\AA}) - \nu_j(7331\textrm{\AA}))]^2 \left.\right\}\,,
\end{eqnarray}
where $w_j$ are the data-set weights (here equal to one except for the
case noted below), $\Delta(\Ddx-\Ssx)$ and $\Delta(\Ppx-\Ddx)$ are the
energy differences (in cm$^{-1}$) between the \Ddn\ and \Ssn\, and the
\Ppn\ and \Ddn\ terms, respectively, $\Delta(\Ddx)$ and $\Delta(\Ppx)$
are the fine structure energy level splitting, and $\nu_j$ are the
observed vacuum wave-numbers of all lines from the $j$ data-sets where
available and corrected for any Doppler shift determined from the
previous fit (nebular sources only).  The Bowen(1960) data was not
included in the $\Delta(\Ppx)$ fitting and its data was given a weight
($w_1=0.5$) for purposes of the $\Delta(\Ppx-\Ddx)$ fit.  The energy
of each term was taken to lie exactly between the fine structure level
energies.

The resultant energy levels, line vacuum wave-numbers, and line air
wavelengths are listed in Table~\ref{table3}.  Listed errors are the
two-sigma formal uncertainties from the correlation matrices of the
various fits, propagated through the equations for line vacuum
wave-numbers as combinations of energy levels/term splittings:
\begin{eqnarray}
\nu(3726\textrm{\AA}) & = & \Delta(\Ddx-\Ssx) + \case{1}{2}\Delta(\Ddx)\,, \\
\nu(3729\textrm{\AA}) & = & \Delta(\Ddx-\Ssx) - \case{1}{2}\Delta(\Ddx)\,, \\
\nu(7319\textrm{\AA}) & = & \Delta(\Ppx-\Ddx) + \case{1}{2}\Delta(\Ppx) + \case{1}{2}\Delta(\Ddx) \,, \\
\nu(7320\textrm{\AA}) & = & \Delta(\Ppx-\Ddx) - \case{1}{2}\Delta(\Ppx) + \case{1}{2}\Delta(\Ddx) \,, \\
\nu(7330\textrm{\AA}) & = & \Delta(\Ppx-\Ddx) + \case{1}{2}\Delta(\Ppx) - \case{1}{2}\Delta(\Ddx) \,, \\ 
\nu(7331\textrm{\AA}) & = & \Delta(\Ppx-\Ddx) - \case{1}{2}\Delta(\Ppx) - \case{1}{2}\Delta(\Ddx) \,,
\end{eqnarray}
where all symbols have the same meanings as above.  The trans-auroral
lines at \lamm2470.2,2470.3\AA, while not observed, were also
calculated from the four fitted energy level and term separations.

\section{Discussions}

The doublet wavelength separation values determined here compare very
favorably with what we consider the most accurately determined
astrophysical values from \citet{BK88} (1.09$\pm$0.02\AA\ for
$\Delta(\Ppx)$ and 10.70$\pm$0.03\AA\ for $\Delta(\Ddx)$, specific
line wavelengths were not reported), while their formal uncertainties
exceed ours.  The \Ddn\ and \Ppn\ term splitting values fall in the
middle of the range of those determined from more recent
experimentally inferred and astrophysically observed values (see
Table~\ref{table1}), although ours should be inherently more accurate
given our direct measurement of the 7320,7330 line wavelengths, as
opposed to Ritz method determinations from more easily observable UV
lines in laboratory experiments.  The measurements made here were not
complicated by blending with other telluric features or significant
thermal broadening of the lines, and thus were not reliant upon
potentially error-prone deblending techniques (e.g.\ DeRobertis et
al.\ 1985).  Our measurements of the \Ppn\ energy splitting, along
with other recent observationally determined values
(Table~\ref{table1}) are consistently and significantly smaller than
those determined from the recent theoretical calculations of
\citet{Z87} (3.0 \cmm\ for their ``A'' configuration, 2.7 \cmm\ for
their ``B'' configuration) and \citet{T02} (2.58 \cmm\ and 2.60 \cmm\
relating to \textit{ab initio} and energy corrected values
respectively), although the \Ddn\ energy splitting values show much
better agreement.  Finally, the application of the new individual
7320,7330 line wavelengths to the results of \citet{B00} significantly
reduces the scatter in velocity residuals with their observed
wavelengths for these lines, from 2.4--8.0 \kms to 2.2--2.7 \kms.
Adoption of the new 7320,7330 wavelengths would lead to an upwards
adjustment in the often-utilized NIST values for the 7320 lines
(7318.92\AA\ and 7319.99\AA) by 0.124 and 0.131\AA\ respectively, and
a one digit increase in the number of significant digits for which the
wavelengths are known.

It appears likely that the splitting value for the \Ppn\ term of
0.796\AA\ determined by \citet{C03} using their imaging Fabry-Perot
interferometer was not that of the 7320 components.  Their
instrument's free spectral range of 0.107\AA\ is almost exactly 1/10
that of the measured 1.077\AA\ \Ppn\ splitting value determined here,
resulting in an unfortunate blending of the lines in their
observations.  It is more likely that the splitting measured was
actually between the stronger \lam7320.121 component and a neighboring
strong \oh line from a different order.  We speculate that the 0.8\AA\
splitting value quoted from \citet{S82} may date back to \citet{M49}
(see Table~\ref{table1}) because their resolution was also
insufficient to separate 7320.  As noted here, \citet{M49} has been
superseded by more recent determinations.  The large differences
between the wavelength and term splitting values determined here, and
other older and/or widely-utilized values, warrant reconsideration of
conclusions reached employing those standards, such as regarding
F-layer physical conditions \citep{S82,C03} and nebular velocity
structure \citep{B00}.
 
Intensities from individual 7320,7330 have been measured in the four
HIRES spectra showing the strongest 7320,7330 lines and least amount
of atomic Fraunhofer absorption from scattered moon or zodiacal light.
The average line intensities normalized to the total 7320,7330 intensity
and standard deviations of the averages are listed in
Table~\ref{table4}.  As the bulk of the atmospheric 7320,7330 emission
arises from altitudes between 250--400 km \citep{C03}, where quenching
of the \Ppn\ term via collisions with atomic O and N$_2$ (of densities
approximating $1\times10^9$ \cmc\ and rate coefficients of
$0.5\times10^{-10}$ \cmc\ s$^{-1}$ and $1.8\times10^{-10}$
respectively; Stephan et al.\ 2003) is more rapid than the term's
radiative lifetime ($\approx$5s), the \oz\ level populations are
thermalized.  Thus, relative intensity measurements of 7320,7330 lines
are related only to spontaneous transition coefficients and
statistical weights.  This provides a check on the accuracy of \oz\
7320,7330 Einstein A values, which are employed in a variety of
astrophysical diagnostics, such as for electron density
\citep{D85,BK88} and temperature \citep{K99}, reddening \citep{D85},
and \oz\ nebular abundances \citep{L00}.

Unfortunately, in most of the HIRES spectra, the 7320,7330 lines were
positioned close to the blue edge of the echelle orders in which they
appeared as a consequence of the original observers' chosen grating
tilts.  Coupled with these lines' intrinsic weakness and confusion of
the continuum level near Fraunhofer absorption features, even in those
spectra with the smallest degree of contamination, the HIRES spectra
are not optimal for intensity measurements of these lines, as
reflected in the large measurement uncertainties.  Still, the
theoretical high-density limit line intensities constructed from major
astrophysical sources of transition coefficients (those listed in
Table~\ref{table4}), are all encompassed within those standard
deviations, and with the possible exception of the \lam7319.044 line,
do compare well with the observed values.  \citet{S79} has previously
reported a 7320/7330 auroral intensity ratio measurement of
1.55$\pm0.05$, which exceeded their own calculation of the ratio's
theoretical limiting value at high density of 1.31 using \citet{SO57}
transition coefficients under maximum quenching conditions (altitude
$<$200 km).  A re-calculation of the theoretical high-density limit
ratio using eqs.(1-6) of \citet{S79} with more recent coefficients
(sources from Table~\ref{table4}) yields values between 1.28 and 1.30,
virtually unchanged from the earlier estimation.  However, our
measured 7320/7330 ratio of 1.3 is consistent with this value and with
the high-density limit theoretical values constructed from
Table~\ref{table4}.

\citet{D85} and \citet{BK88} have proposed the use of the
\lam7320.121/\lam7319.044 intensity ratio as an electron density
diagnostic in the $\approx10^5$--$10^9$\cmc\ density regime.  The
diagnostic takes advantage of the increase in the theoretical
intensity ratio, from a value of about 3.0 below $10^5$ \cmc, to a
high-density value of 3.8 (as calculated from Table~\ref{table4}) at
$10^9$ \cmc\ and above, as electron collisional excitation from the
\Ddn\ term and de-excitation increasingly contribute to the \Ppn\
levels' populations.  The average value of the
\lam7320.121/\lam7319.044 intensity ratio measured from the HIRES
spectra is 3.1$\pm$0.9, while values of 3.3 and 3.5 are measured from
the blended profile of the 7320 doublet in two lower resolution Keck
II/ESI spectra in our possession, all of which are lower than the
composite high-density ratio value of 3.8.  A lower value of the
high-density limit intensity ratio, as is suggested from the HIRES and
ESI spectra measurements, would reduce the utility of the diagnostic.

A stronger than predicted \lam7319.044 intensity relative to the total
multiplet intensity, as is suggested by the HIRES data, could be
behind the observed smaller than predicted value of the
\lam7320.121/\lam7319.044 intensity ratio.  However, the line's large
intensity measurement uncertainty in the HIRES spectra, echoed into
the large uncertainty in the measured intensity ratio, suggests that
the difference between the observed and theoretical high-density
values of the ratio is probably not statistically significant.  The
ESI spectra ratio values are closer to composite theoretical
high-density ratio value of 3.8, and systemic errors introduced in the
deblending of a 7320 profile composed of unequal intensity components
could account for much of the difference.  Significant contamination
of the 7320,7330 line intensities from the N$_2$ 1P(5-3) band is ruled
out since the band head (at 7387.2\AA) and strongest
features, as predicted from a DIATOM \citep{H94} simulation of the
band's intensity at a typical mesospheric temperature of 200K, are
absent from any of the utilized spectra.

Curiously, the \lam7320.121/\lam7319.044 intensity ratios measured
from the IC 418 and Orion Nebula spectra of 2.7 and 2.9 respectively,
as calculated from Table~\ref{table4}, are slightly below the 3.0
low-density value, even though both objects have electron densities of
$\approx10^4$ \cmc.  This may indicate that additional processes other
than electron collisional excitation may be populating the \Ppn\
levels not according to their statistical weights.  For example,
electron recombination of O$^{+2}$ with subsequent cascade
\citep{L00}, or photoionization of neutral O \citep{F81}, are two
processes that may also affect the nebular 7320,7330 line intensities.
  
In summary, the HIRES data generally supports the accuracy of
widely-used astrophysical spontaneous transition coefficients, with
some lingering problems which the present data-sets' large scatter and
small numbers are unable to resolve.

\section{Conclusions}

We have used high resolution HIRES sky spectra to accurately determine
the wavelengths and relative intensities of the 7320,7330 doublet
lines, measured directly for the first time from completely resolved
profiles in the airglow.  These measurements are extremely accurate
given the nature and advantages of using sky spectra for such
measurements as opposed to deducing the wavelengths from energy levels
established by more easily observable transitions, or from nebular
lines wavelengths corrected for proper and internal kinematics.  The
new wavelengths determined here differ by as much as 0.131\AA\ from
the widely-utilized NIST values, and the aeronomically-important 7320
doublet splitting value measured here differs by 0.277\AA\ from the
``accepted'' value of 0.8\AA\ still employed as recently as
\citet{C03}.  The magnitudes of these differences suggest that the
interpretation of observations made by comparing against earlier
standards may need to be revisited.  The observed fluxes for
individual lines agree within their sample scatter, with theoretical
values calculated under the high density limit which should prevail in
the region of the atmosphere in which they are formed.

\acknowledgements 

This paper is based on observations made at the W.M.\ Keck
Observatory, which is operated jointly by the California Institute of
Technology and the University of California.  We thank Professor
Jack Baldwin for allowing us to use his spectra.  This work was
supported by grants from the NSF Aeronomy Program and the NASA Office
of Space Science.

\clearpage
\clearpage

\begin{deluxetable}{lcccccccc}
\tablecolumns{9}
\tabletypesize{\footnotesize}
\tablecaption{Selected Sources of Energy Levels for \oz\ \tu \label{table1}}
\tablehead{
         & \multicolumn{8}{c}{Level Energy (cm$^{-1}$)} \\ \cline{2-9}
 &Moore&Bowen&Fawcett&DO\tablenotemark{(a)}&Erikkson&MKM\tablenotemark{(b)}&TFF\tablenotemark{(c)}&This Work\\      
\multicolumn{1}{c}{Term} &(1949)&(1960)&(1975)&(1985)&(1987)&(1993)&(2002)& 
}
\startdata
\Ss      & \phn\phn\phn\phn0.0    &\phn\phn\phn\phn0.0    & \phn\phn\phn\phn0.0  &\phn\phn\phn\phn0.0  & \phn\phn\phn\phn0.00 & \phn\phn\phn\phn0.00 & \phn\phn\phn\phn0.00 & \phn\phn\phn\phn0.00 \\
\Dd{5/2}&  26808.4  & 26810.7 &  26810.7  &  26810.5  & 26810.52 & 26810.55 & 26810.73 & 26810.76 \\    
\Dd{3/2}&  26829.4  & 26830.5 &  26830.2  &  26830.6  & 26830.57 & 26830.57 & 26830.45 & 26830.57 \\
\Pp{3/2}&  40466.9  & \raisebox{-1.5ex}[0cm][0cm]{$\left\}\,\,40468.3\,\,\right\{$} & 40467.5  &  40468.1  & 40467.69 & 40468.01 & 40468.36 & 40467.97 \\
\Pp{1/2}&  40468.4  &         &  40468.6  &  40470.1  & 40469.69 & 40470.00 & 40470.96 & 40469.98 
\enddata
\tablenotetext{(a)}{\citet{D85}}
\tablenotetext{(b)}{\citet{M93}, Source of NIST \oz\ \tu wavelengths}
\tablenotetext{(c)}{\citet{T02}, Energy adjusted values}
\end{deluxetable}

\clearpage

\begin{deluxetable}{lcccrcclc}
\tabletypesize{\footnotesize}
\tablecaption{HIRES Spectra Observing Journal \label{table2}}
\tablehead{
\multicolumn{1}{c}{Date} &
\colhead{UT} &
\multicolumn{1}{c}{Exposure} &   
\colhead{Alt} &
\multicolumn{1}{c}{Azi} &     
\colhead{RA} &            
\colhead{DEC} &
\multicolumn{1}{c}{Slit} &
\colhead{Observers} \\
\multicolumn{1}{c}{(UT)} & & \multicolumn{1}{c}{(sec)} & & & (J2000) & (J2000) & \multicolumn{1}{c}{(\arcsec)} &  
}
\startdata 
1993 Nov 15 &  04:33 & \phn150  &   80.97  &   4.50  & 21:51:11.1  & +28:51:53.5  &  0.861$\times$14.0 & \tablenotemark{(a)} \\
1993 Nov 15 &  05:05 & 3000  &   53.94  & 182.41  & 22:15:27.2  & $-$16:11:33.0  &  0.861$\times$14.0 & \tablenotemark{(a)} \\
1996 Aug 07 &  05:49 & \phn300  &   35.70  & 277.81  & 12:41:51.9  & +17:31:22.3  &  0.861$\times$14.0 & \tablenotemark{(b)} \\
1999 Jun 14 &  05:39 & \phn600  &   58.42  & 324.19  & 11:05:30.9  & +43:31:13.5  &  0.861$\times$7.0 & \tablenotemark{(c)} \\
1999 Jun 14 &  05:54 & 1200  &   54.92  & 313.35  & 10:47:12.7  & +40:26:46.4  &  0.861$\times$7.0 & \tablenotemark{(c)} \\
1999 Jun 14 &  06:17 & 2400  &   50.97  & 310.57  & 10:47:13.9  & +40:26:53,1  &  0.861$\times$7.0 & \tablenotemark{(c)} \\
1999 Jun 14 &  14:13 & 2400  &   58.61  &  84.75  & 23:34:39.5  & +19:33:04.5  &  0.861$\times$7.0 & \tablenotemark{(c)} \\
1999 Jun 15 &  05:40 & \phn600  &   57.73  & 323.12  & 11:05:31.4  & +43:31:20.2  &  0.861$\times$7.0 & \tablenotemark{(c)} \\
1999 Jun 15 &  05:54 & 3000  &   54.25  & 312.81  & 10:47:14.1  & +40:26:53.7  &  0.861$\times$7.0 & \tablenotemark{(c)} \\
1999 Jun 15 &  14:21 & 1800  &   75.12  &  91.88  & 22:35:49.1  & +18:40:30.9  &  0.861$\times$7.0 & \tablenotemark{(c)} \\
\enddata
\tablenotetext{(a)}{W.L.W. Sargent, J.K. McCarthy}
\tablenotetext{(b)}{M. Rauch, L. Lu}
\tablenotetext{(c)}{I.N. Reid, J.D. Kirkpatrick, A.J. Burgasser, J. Liebert}
\end{deluxetable}

\clearpage

\begin{deluxetable}{lcccc}
\tablecolumns{5}
\tabletypesize{\footnotesize}
\tablewidth{4.6in}
\tablecaption{Energy Levels, Line Wavelengths, and Term Splittings \label{table3}}
\tablehead{
 &  Level & Wavelength (air) & NIST & $\Delta$\tablenotemark{(a)} \\
Transition & (cm$^{-1}$) & (\AA) & (\AA) & (\AA) 
}
\startdata
\Ss--\Pp{1/2}     &  40469.98\phn$\pm$0.08\phn & 2470.220$\pm$0.005$\,$ & 2470.219 & +0.001 \\
\Ss--\Pp{3/2}     &  40467.97\phn$\pm$0.08\phn & 2470.343$\pm$0.005$\,$ & 2470.341 & +0.002 \\
\Ss--\Dd{3/2}     &  26830.57\phn$\pm$0.08\phn & 3726.032$\pm$0.011$\,$ & 3726.032 & \nodata \\
\Ss--\Dd{5/2}     &  26810.76\phn$\pm$0.08\phn & 3728.785$\pm$0.011$\,$ & 3728.815 & -0.030 \\
\Dd{5/2}--\Pp{1/2}&  13659.223$\pm$0.007       & 7319.044$\pm$0.004$\,$ & 7318.92\phn & +0.124 \\
\Dd{3/2}--\Pp{1/2}&  13657.213$\pm$0.007       & 7320.121$\pm$0.004$\,$ & 7319.99\phn & +0.131 \\
\Dd{5/2}--\Pp{3/2}&  13639.413$\pm$0.007       & 7329.675$\pm$0.004$\,$ & 7329.67\phn & +0.005 \\
\Dd{3/2}--\Pp{3/2}&  13637.403$\pm$0.007       & 7330.755$\pm$0.004$\,$ & 7330.73\phn & +0.025 \\
$\Delta(\Ppx)$\tablenotemark{(b)} &  \phn\phn\phn\phn2.010$\pm$0.005       &\phn\phn\phn1.077$\pm$0.003 & \phn\phn\phn1.07\phn & +0.007 \\
\hspace*{2.6pc}\tablenotemark{(c)}  &                            & \phn\phn\phn1.080$\pm$0.003 & \phn\phn\phn1.06\phn & +0.020 \\
$\Delta(\Ddx)$\tablenotemark{(d)} &  \phn\phn\phn19.810$\pm$0.006       & \phn\phn10.630$\pm$0.003 & \phn\phn10.75\phn & -0.120 \\
\hspace*{2.8pc}\tablenotemark{(e)}  &                            &  \phn\phn 10.633$\pm$0.003 & \phn\phn10.74\phn & -0.107 \\
\enddata
\tablenotetext{(a)}{in the sense of correction to NIST to align with present values}
\tablenotetext{(b)}{calculated from \lam(7320\AA)-\lam(7319\AA)}
\tablenotetext{(c)}{calculated from \lam(7331\AA)-\lam(7330\AA)}
\tablenotetext{(d)}{calculated from \lam(7330\AA)-\lam(7319\AA)}
\tablenotetext{(e)}{calculated from \lam(7331\AA)-\lam(7320\AA)}
\end{deluxetable}

\clearpage

\begin{deluxetable}{lllll}
\tablecolumns{5}
\tablecaption{Relative Strengths of 7320,7330 Lines \tablenotemark{(a)} \label{table4}}
\tablehead{
 & \multicolumn{4}{c}{Intensity} \\ \cline{2-5}
Reference &  \multicolumn{1}{c}{\lam7319.044} &  \multicolumn{1}{c}{\lam7320.121} &  \multicolumn{1}{c}{\lam7329.675} &  \multicolumn{1}{c}{\lam7330.755}
}
\startdata
%\citet{Z82} & 0.12 & 0.45 & 0.20 & 0.24 \\
\citet{GFF84}\tablenotemark{(b)} & 0.12 & 0.45 & 0.20 & 0.24 \\
\citet{Z87}\tablenotemark{(c)} & 0.12 & 0.45 & 0.20 & 0.24 \\
\citet{WFD96}\tablenotemark{(d)} & 0.12 & 0.45 & 0.20 & 0.24 \\
\citet{B00} & 0.14 & 0.41 & 0.22 & 0.23 \\
\citet{S03} & 0.15 & 0.40 & 0.23 & 0.22 \\
This Paper & 0.14$\pm$0.05 & 0.43$\pm$0.03 & 0.20$\pm$0.04 & 0.23$\pm$0.03  
\enddata
\tablenotetext{(a)}{normalized to sum of all multiplet lines, assuming statistically populated levels}
\tablenotetext{(b)}{same within precision for both configuration calculations}
\tablenotetext{(c)}{configuration sets ``A'' and ``B'' with theoretical energy correction, relativistic M1 values}
\tablenotetext{(d)}{source of NIST \oz\ values, arithmetic average of \citet{Z87} and \citet{GFF84} values, the later modified with newer experimental wavelength data} 
\end{deluxetable}

\end{document}